\documentclass[twocolumn,showpacs,aps,prb]{revtex4}
\usepackage{amssymb,amsmath}
\usepackage[dvips]{graphicx}

\usepackage[dvipdfm,letterpaper]{hyperref}

\begin{document}

\title{Infinite randomness and ``quantum'' Griffiths effects in a classical system:
       the randomly layered Heisenberg magnet}

\author{Priyanka Mohan}
\author{Rajesh Narayanan}
\affiliation{Department of Physics, Indian Institute of Technology Madras, Chennai 600036, India}
\author{Thomas Vojta}
\affiliation{Department of Physics, Missouri University of Science and Technology, Rolla,
MO 65409, USA}

\begin{abstract}
We investigate the phase transition in a three-dimensional classical Heisenberg
magnet with planar defects, i.e., disorder perfectly correlated in two dimensions.
By applying a strong-disorder renormalization group, we show that the critical
point has exotic infinite-randomness character. It is accompanied by strong
power-law Griffiths singularities. We compute various thermodynamic observables
paying particular attention to finite-size effects relevant for an experimental
verification of our theory. We also study the critical dynamics within a Langevin
equation approach and find it extremely slow. At the critical point, the
autocorrelation function decays only logarithmically with time while it follows
a nonuniversal power-law in the Griffiths phase.

\end{abstract}


\date{\today}
\maketitle

\section{Introduction}
\label{Sec:Intro}
The influence of impurities, defects, or other types of quenched disorder on the
properties of phase transitions has aroused the interest of physicists for more than
three decades (see Ref.\ \onlinecite{Grinstein85} for an overview of some of the early
work). Recently, this field has reattracted considerable attention as it has become clear
that disorder effects are generically much stronger at zero-temperature quantum phase
transitions than at classical thermal phase transitions. This leads to unconventional
phenomena such as power-law quantum Griffiths
singularities,\cite{ThillHuse95,GuoBhattHuse96,RiegerYoung96} infinite-randomness
critical points with exponential rather than power-law scaling,\cite{Fisher92,Fisher95}
or even smeared phase transitions.\cite{Vojta03a,HoyosVojta08} A recent review of part of
this physics can be found in Ref.\ \onlinecite{Vojta06}.

The main reason for the enhanced disorder effects at quantum phase transitions
is that the disorder is perfectly correlated in \emph{imaginary time} direction.
Because imaginary time acts as an extra dimension at a quantum phase transition
(and becomes infinitely extended at zero temperature), one is effectively dealing
with defects that are ``infinitely  large'' in this extra dimension. Thus, they
are much harder to average out than conventional finite-size defects.

This implies that similarly strong effects can be expected at a classical thermal phase
transition if the disorder is perfectly correlated in one or more \emph{space}
dimensions. Indeed, it has been known for a long-time that the McCoy-Wu model, a
classical two-dimensional Ising model with disorder perfectly correlated in one of the
two dimensions, exhibits an unusual phase transition. In a series of
papers,\cite{McCoyWu68,McCoyWu68a,McCoyWu69,McCoy69} McCoy and Wu developed a
transfer-matrix approach to this model and showed that the specific heat is smooth across
the ferromagnetic phase transition while the susceptibility is infinite over an entire
temperature range. Fisher\cite{Fisher92,Fisher95} later achieved an essentially complete
understanding of this transition by means of a strong-disorder renormalization group
(using the equivalence between the McCoy-Wu model and the one-dimensional random
transverse-field Ising chain). He found that the critical point is of infinite-randomness
type, and it is accompanied by strong power-law Griffiths singularities. Largely due to the
fact that the McCoy-Wu model is difficult to realize in nature, these predictions have
(to the best of our knowledge) not been experimentally verified yet.

In this paper, we present another classical system exhibiting an exotic
infinite-randomness critical point, viz., a randomly layered three-dimensional
Heisenberg magnet. This system is more easily realizable in experiment than
the McCoy-Wu model as it
can be produced by depositing random layers of two different ferromagnetic materials.
Moreover, because of its three-dimensional character, it permits bulk thermodynamic
measurements. We investigate the phase transition in this model by means of a
strong-disorder renormalization group which allows us to determine the critical
behavior exactly.

Our paper is organized as follows: In Sec.\ \ref{Sec:Model}, we introduce the
randomly layered Heisenberg model and give heuristic arguments for the strong
disorder effects. In Sec.\ \ref{Sec:SDRG} we explain our theoretical approach.
Results on the
thermodynamics are given in Sec.\ \ref{Sec:TD}, while the experimentally important
finite-size effects are discussed in Sec.\ \ref{Sec:FSE}. Section \ref{Sec:Dynamics}
is devoted to the dynamical behavior at the phase transition. We conclude in Sec.\
\ref{Sec:Conclusions}.

\section{Randomly layered Heisenberg model}
\label{Sec:Model}

\begin{figure}
\includegraphics[width=8cm]{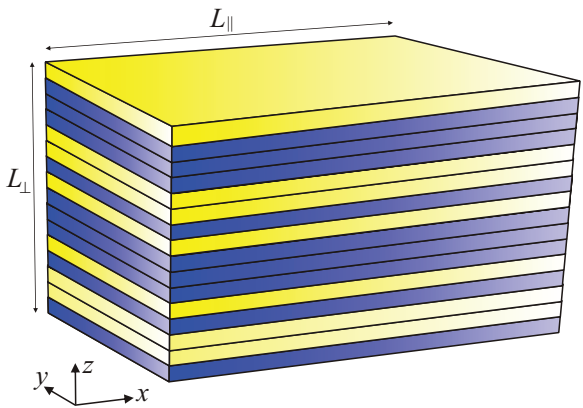}
\caption{Schematic of the layered magnet: layers of two different ferromagnetic materials
are arranged in a random sequence.}
\label{Fig:layeredmagnet}
\end{figure}

We consider a three-dimensional Heisenberg ferromagnet consisting of a random sequence of
layers made of two different ferromagnetic materials, as sketched in Fig.\
\ref{Fig:layeredmagnet}.
This system can be modelled by a classical Heisenberg Hamiltonian on a cubic lattice
given by
\begin{equation}
H = - \sum_{\mathbf{r}} J^{\parallel}_z \, (\mathbf{S}_{\mathbf{r}} \cdot \mathbf{S}_{\mathbf{r}+\hat{\mathbf{x}}}
                                        +\mathbf{S}_{\mathbf{r}} \cdot \mathbf{S}_{\mathbf{r}+\hat{\mathbf{y}}} )
    - \sum_{\mathbf{r}} J^{\perp}_z \, \mathbf{S}_{\mathbf{r}} \cdot\mathbf{S}_{\mathbf{r}+\hat{\mathbf{z}}}
    .
\label{Eq:Hamiltonian}
\end{equation}
Here, $\mathbf{S}_{\mathbf{r}}$ is a three-component unit vector on lattice site
$\mathbf{r}$, and  $\hat{\mathbf{x}}$, $\hat{\mathbf{y}}$, and $\hat{\mathbf{z}}$ are
the unit vectors in the coordinate directions. The exchange interactions within
the layers, $J^{\parallel}_z$, and between the layers, $J^{\perp}_z$, are both positive
and independent random functions of the perpendicular coordinate $z$.

To develop a heuristic understanding of the randomly layered Heisenberg model, we first
consider the case of all $J^{\perp}_z$ being identical, $J^{\perp}_z \equiv J^{\perp}$,
while the $J^{\parallel}_z$ are drawn from a binary probability distribution
\begin{equation}
P(J^{\parallel})=(1-p)\, \delta(J^{\parallel} - J_u) + p\, \delta(J^{\parallel} - J_l)
\label{Eq:binary_distrib}
\end{equation}
with $J_u > J_l$. Here, $p$ is the concentration of the ``weak'' layers while $1-p$ is
the concentration of the ``strong'' layers. More general distributions will be considered
in the next section.

\begin{figure}
\includegraphics[width=8cm, clip]{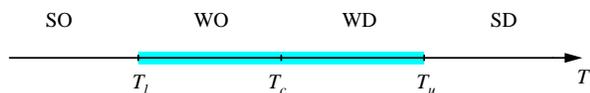}
\caption{Schematic phase diagram of the randomly layered Heisenberg magnet
(\ref{Eq:Hamiltonian}). SD and SO denote the conventional strongly disordered
and strongly ordered phases, respectively. WD and WO are the weakly disordered
and ordered Griffiths phases. $T_c$ is the critical temperature while $T_u$ and $T_l$
mark the boundaries of the Griffiths phase.}
\label{Fig:pd}
\end{figure}

Let us now discuss the behavior of the model (\ref{Eq:Hamiltonian}) qualitatively
(see Fig.\ \ref{Fig:pd}).
At sufficiently high
temperatures, the system will be in a conventional (strongly disordered) paramagnetic
phase with a finite magnetic susceptibility which increases upon lowering the temperature.
Below a temperature $T_u$ (which is the transition temperature of a hypothetical system containing
strong layers only, $J^{\parallel}_z \equiv J_u$), rare thick slabs of strong layers
develop local order while the bulk system is still nonmagnetic. This is the weakly disordered
Griffiths phase. The Griffiths phase continues below the actual critical temperature
$T_c$ down to a temperature $T_l$ (which is the transition temperature of a hypothetical
system containing weak layers only, $J^{\parallel}_z \equiv J_l$). In the weakly ordered
Griffiths phase, bulk magnetism coexists with locally nonmagnetic slabs. Finally, below
$T_l$, the system is in a conventional (strongly ordered) ferromagnetic phase.

To estimate the strength of the Griffiths singularities in this system, we need to
compare the probability of finding a thick slab of strong layers with the contribution
such a slab can make to thermodynamic quantities such as the susceptibility. Simple
combinatorics gives the probability for finding a slab of $L_{RR}$ consecutive strong layers
to be
\begin{equation}
w(L_{RR}) \sim (1-p)^{L_{RR}} = e^{-\tilde p L_{RR} }
\label{Eq:w(L_RR)}
\end{equation}
with $\tilde p = -\ln(1-p)$. Each such slab is equivalent to a two-dimensional
Heisenberg model with an effective exchange interaction $L_{RR} J_u$. Because the two-dimensional
Heisenberg model is exactly at its lower critical dimension, the susceptibility of the
slab increases exponentially with the effective interaction,\cite{VojtaSchmalian05,Vojta06}
\begin{equation}
\chi(L_{RR}) \sim e^{b L_{RR}}
\label{Eq:chi(L_RR)}
\end{equation}
where $b$ increases with decreasing temperature.
The same result also follows from a renormalization group analysis of the corresponding
nonlinear sigma model at its low-temperature fixed point \cite{Toner_unpublished05} or
from an explicit large-$N$ calculation (as shown in the next section).

Thus, the exponential \emph{decrease} of the rare region probability $w(L_{RR})$ with size
$L_{RR}$ is compensated by an exponential \emph{increase} in the contribution it makes to
the susceptibility. The total rare region susceptibility in the weakly disordered Griffiths
phase is obtained by simply summing over the contributions of the individual rare
regions. Up to pre-exponential factors, this yields
\begin{equation}
\chi_{RR} = \int dL_{RR} e^{(b-\tilde p) L_{RR}}~.
\label{Eq:RR_susceptibility}
\end{equation}
The total rare region susceptibility thus diverges once $b$ becomes larger than $\tilde
p$. Other observables can be discussed along the same lines.
Equations (\ref{Eq:w(L_RR)}), (\ref{Eq:chi(L_RR)}), and (\ref{Eq:RR_susceptibility}) are
analogous to the corresponding relations for the McCoy-Wu
model \cite{McCoyWu68,McCoyWu68a,McCoyWu69,McCoy69}
or (after quantum-to-classical mapping) to those of the random transverse-field Ising
model. \cite{ThillHuse95,GuoBhattHuse96,RiegerYoung96}
This suggests that the phase transition in our model displays unconventional behavior.
In the next section we investigate this question in detail by means of a
renormalization group method.

\section{Strong-disorder renormalization group}
\label{Sec:SDRG}

In this section we study the ferromagnetic phase transition of the randomly layered
Heisenberg model by means of a strong-disorder renormalization
group.\cite{MaDasguptaHu79,IgloiMonthus05} Our implementation
of this method follows a recent study of dissipative quantum phase transitions.
\cite{HoyosKotabageVojta07,VojtaKotabageHoyos09} We therefore only outline the major
steps of the calculation, details can be found in Ref.\
\onlinecite{VojtaKotabageHoyos09}.

\subsection{Order parameter field theory}
\label{Subsec:LGW}

Our starting point is a Landau-Ginzburg-Wilson (LGW) order parameter field theory
for an $N$-component order parameter $\phi(\mathbf{r})$. In the absence of disorder, the free energy functional reads
\begin{equation}
S = \int d^3 r \left[ \delta_0 \, \phi^2(\mathbf{r}) + \gamma_0^2 (\partial_\mathbf{r}
   \phi(\mathbf{r}))^2  + u\, \phi^4(\mathbf{r}) \right]~.
\label{Eq:LGW}
\end{equation}
Here, $\delta_0$ is the bare distance from criticality, $\gamma_0$ is the bare
length scale, and $u$ is the $\phi^4$ coefficient. In the presence of our layered disorder,
$\delta_0$, $\gamma_0$, and $u$ become random functions of the $z$-coordinate (the coordinate
perpendicular to the layers) and the derivative term will generally be
anisotropic. In order to apply the real-space-based strong-disorder
renormalization group, we discretize the continuum LGW theory (\ref{Eq:LGW})
in the $z$-direction, but not in the $xy$-plane.
For simplicity, we first consider the large-$N$ limit of our
LGW theory which allows us to perform all of the following
calculations explicitly. We will later show that the resulting critical point is the
same for all $N>2$ including the physically relevant Heisenberg case $N=3$.
The discrete large-$N$ order parameter field
theory reads
\begin{equation}
S = \sum_{z,\mathbf{q}} (\delta_z +\lambda_z + \gamma_z^2 \mathbf{q}^2)|\phi_z(\mathbf{q})|^2
 - \sum_{z,\mathbf{q}} J_z^\perp \phi_z(\mathbf{q}) \phi_{z+1}(\mathbf{-q})
\label{Eq:LGW_discrete}
\end{equation}
where $\mathbf{q}$ is a two-component vector describing the $xy$-momentum. The Lagrange
multipliers $\lambda_z$ enforce the large-$N$ constraints
$\langle(\phi_{z}^{(k)})^{2}\rangle=1$ for the $k$-th order parameter component in layer
$z$; they have to be determined self-consistently. The renormalized local distance from
criticality in layer $z$ is given by $\epsilon_{z}=\delta_{z}+\lambda_{z}$. In the
disordered phase, \emph{all} $\epsilon_z>0$. For the case of a single layer, the LGW
theory (\ref{Eq:LGW_discrete}) can be solved immediately, giving
\begin{equation}
\epsilon_z = \gamma_z^2 \Lambda^2 e^{-4\pi \gamma_z^2 / a^2}
\label{Eq:single_layer}
\end{equation}
with $\Lambda$ being a momentum cutoff and $a$ the lattice constant.

\subsection{Recursion relations}
\label{Subsec:Recursions}

The basic idea of the strong-disorder renormalization group is to successively
integrate out local high-energy degrees of freedom. In the LGW theory (\ref{Eq:LGW_discrete}),
the competing local couplings are the local distances from criticality $\epsilon_z$ and the interactions $J^\perp_z$.
In the bare theory, they are independent random variables with distributions $R_0(\epsilon)$ and $P_0(J^\perp)$, respectively.
The method relies on these distributions being broad and becomes exact in the limit of
infinitely broad distributions. We will verify this condition a-posteriori.

In each renormalization group
step, we choose the largest local coupling $\Omega=\max\{\epsilon_z,J_{z}^\perp\}$.
If it is a distance from criticality, say $\epsilon_{2}$, the unperturbed part of the
free energy
is $S_{0}=\sum_{\mathbf{q}}(\epsilon_{2}+\gamma_2^2 \mathbf{q}^2)|\phi_{2}(\mathbf{q})|^{2}$.
The coupling of $\phi_{2}$ to the neighboring layers, $S_{1}=-\sum_{\mathbf{q}}
[J_{1}^\perp \phi_{1}(\mathbf{q})\phi_{2}(-\mathbf{q}) + J_{2}^\perp \phi_{2}(\mathbf{q})\phi_{3}(-\mathbf{q})]$,
is treated perturbatively. Keeping only the leading long-wavelength
terms that arise in 2nd order of the cumulant expansion, we obtain
new interactions $\tilde{S}=-\sum_{\mathbf{q}}\tilde{J}_{1}^\perp \phi_{1}(\mathbf{q}) \phi_{3}(-\mathbf{q})$
with
\begin{equation}
\tilde{J}_{1}^\perp= \frac{J_{1}^\perp J_{2}^\perp}{\epsilon_{2}}\,.
\label{eq:J-tilde}
\end{equation}
At the end of the renormalization group step, $\phi_{2}$ is dropped from the action.

If the largest local energy is an interaction, say $J_{2}^\perp$, we solve the two-layer
problem
$S_{0}=\sum_{\mathbf{q}}\sum_{z=2,3}(\epsilon_{z}+\gamma_{z}^2 \mathbf{q}^2)|\phi_{z}(\mathbf{q})|^{2}
-\sum_{\mathbf{q}}J_{2}^\perp\phi_{2}(\mathbf{q})\phi_{3}(-\mathbf{q})$
exactly while treating the interactions with the neighboring layers as perturbations.
For $J_{2}^\perp \gg\epsilon_{2},\,\epsilon_{3}$, the two fields $\phi_{2}$ and $\phi_{3}$ are
essentially parallel; thus they can be replaced by a single field $\tilde{\phi}_{2}$ with
an effective renormalized free energy functional
$\tilde{S}=\sum_{\mathbf{q}}(\tilde{\epsilon_{2}}+\tilde{\gamma}_{2}^2 \mathbf{q}^2)|\tilde{\phi_{2}}(\mathbf{q})|^{2}$.

After a straightforward but somewhat lengthy calculation,\cite{VojtaKotabageHoyos09}
the effective distance from criticality of the combined layer comes out to be
\begin{equation}
\tilde{\epsilon}_{2}=2\frac{\epsilon_{2}\epsilon_{3}}{J_{2}^\perp}\,,
\label{eq:e-tilde}
\end{equation}
while the length scale parameter renormalizes as
$\tilde{\gamma_{2}^2}=\gamma_{2}^2+\gamma_{3}^2$.
The new field represents a layer with effective moment per site
\begin{equation}
\tilde{\mu}_{2}=\mu_{2}+\mu_{3}\,.
\label{eq:moment}
\end{equation}
The interactions of the combined layer with the neighboring layers are not
renormalized.
The net result of the renormalization group step is the elimination of one layer and the
reduction of the energy scale $\Omega$.

The structure of the renormalization group recursion
relations (\ref{eq:J-tilde}-\ref{eq:moment})
is identical to those of the one-dimensional random transverse-field
Ising model\cite{Fisher92,Fisher95} as well as the dissipative quantum rotor
model.\cite{HoyosKotabageVojta07,VojtaKotabageHoyos09}
Consequently (and somewhat surprisingly), the \emph{thermal} phase transition in our randomly-layered
classical \emph{three-dimensional} Heisenberg model belongs to the same universality class as
the \emph{quantum} phase transitions in the \emph{one-dimensional} random transverse-field
Ising model and the dissipative quantum rotor chain.

At first glance, this result seems to suggest that crucial system characteristics such as
order parameter symmetry and dimensionality are rendered unimportant by the
strong-disorder renormalization group. However, the physics turns out to be more subtle.
The fact that our randomly layered Heisenberg model and the random transverse-field Ising
chain are in the same universality class is due to a nontrivial interplay between the
order parameter symmetry and the defect dimensionality. We will discuss this point in
more detail in Sec.\ \ref{Sec:Conclusions} in the context of a general classification of
phase transitions in the presence of disorder.

\subsection{Fixed points}
\label{Subsec:Fixed points}

The renormalization group step outlined in the last subsection
does not change the lattice topology because we remove a full layer in each
step. Moreover, the surviving $\epsilon$ and $J^\perp$ remain statistically independent.
The theory can therefore be formulated in terms of individual probability distributions
$P(J^\perp)$ and $R(\epsilon)$. Fisher derived flow equations for these distributions
and solved them analytically.\cite{Fisher92,Fisher95} They have three three kinds
of nontrivial fixed points representing the weakly ordered and disordered Griffiths
phases as well as the critical point in-between. At the critical fixed point,
the relative width of the distributions $P(J^\perp)$ and $R(\epsilon)$ diverges, justifying the method and
giving the critical point its name, \emph{infinite-randomness} critical point.

The critical behavior is characterized by three exponents, $\nu=2$, $\psi=1/2$, and
$\phi=(1+\sqrt{5})/2$. The exponent $\nu$ controls how the perpendicular correlation
length $\xi_\perp$ diverges as the critical point is approached
\begin{equation}
\xi_\perp \sim |\delta|^{-\nu}~.
\label{eq:nu}
\end{equation}
$\xi_\perp$ characterizes the spatial correlations perpendicular to the layers
(in $z$-direction). $\delta$ is the fully renormalized distance from the critical
point; it is given by $\delta \sim [\ln(\epsilon) -\ln(J^\perp)]_0$ in terms of the
bare variables ($[\cdot]_0$ denotes the average over the bare disorder distributions.)

The exponent $\psi$ (which is sometimes called the tunneling exponent because of its
meaning in the quantum problem of Refs.\ \onlinecite{Fisher92,Fisher95})
relates the perpendicular correlation length $\xi_\perp$ and the
correlation length $\xi_\parallel$ within the layers. The scaling is highly anisotropic,
\begin{equation}
\ln(\xi_\parallel/a) \sim \xi_\perp^\psi~.
\label{eq:psi}
\end{equation}
$\psi$ also controls the density $n_\Omega$ of layers surviving at energy scale $\Omega$
in the renormalization procedure. The scaling form of this variable is given
by\cite{VojtaKotabageHoyos09}
\begin{equation}
n_\Omega(\delta) = [\ln(\Omega_I/\Omega)]^{-1/\psi}
X_n\left[\delta^{\nu\psi}\ln(\Omega_I/\Omega)\right]~,
\label{eq:n_scaling}
\end{equation}
where $\Omega_I$ is a constant of the order of the initial (bare) value of $\Omega$. The
scaling function behaves as $X_n(0) =$ const and $X_n(y\to\infty) \sim y^{1/\psi}e^{-cy}$
where $c$ is a constant. As a result, the layer density decreases as  $n_\Omega
\sim[\ln(\Omega_I/\Omega)]^{-1/\psi}$ at criticality while it behaves as $n_\Omega \sim
\delta^{\nu} \Omega^{1/z}$ in the disordered Griffiths phase ($\delta>0$). The
nonuniversal exponent $z$ varies as $z\sim \delta^{-\nu\psi}$ in the Griffiths phase.

The exponent $\phi$ determines how the typical moment $\mu_\Omega$ per site
 of a surviving layer depends on the energy scale $\Omega$.
The scaling form of $\mu_\Omega$ reads
\begin{equation}
\mu_\Omega(\delta) = [\ln(\Omega_I/\Omega)]^{\phi}
X_\mu\left[\delta^{\nu\psi}\ln(\Omega_I/\Omega)\right]~.
\label{eq:mu_scaling}
\end{equation}
The scaling function behaves as $X_\mu(0) =$ const and $X_\mu(y\to\infty)
\sim y^{1-\phi}$. Thus, at criticality the typical moment increases as
$\mu_\Omega \sim [\ln(\Omega_I/\Omega)]^{\phi}$ while it behaves as
$\mu_\Omega \sim \delta^{\nu\psi(1-\phi)} \ln(\Omega_I/\Omega)$
in the disordered Griffiths phase.

\section{Thermodynamics}
\label{Sec:TD}

The overall strategy \cite{Fisher95} for computing the behavior of thermodynamic observables
consists in running the strong-disorder renormalization group from the initial energy
scale $\Omega_I$ down to the energy scale set by an external perturbation such as a
magnetic field. The high-energy degrees of freedom eliminated in this way
do not make significant contributions to the long-wavelength physics.
The surviving layers are very weakly coupled and can be treated as
independent. In this section we show that the resulting thermodynamic behavior
of our system is similar to that at an infinite-randomness quantum critical point.
However, there are significant differences due to the fact that we are dealing with
a thermal (classical) phase transition.

\subsection{Single-layer results}

We start by considering a single layer with effective moment $\mu$ per site
in an external magnetic field (i.e., this layer is the result of combining
$\mu$ original layers during the renormalization group). The free energy
functional is given by
\begin{equation}
S_l = \sum_{\mathbf{q}} (\delta +\lambda + \gamma^2 \mathbf{q}^2)|\phi(\mathbf{q})|^2
 -  \mu \sum_{\mathbf{q}}h(\mathbf{q}) \phi(\mathbf{-q})
\label{Eq:LGW_single}
\end{equation}
where $h(\mathbf{q})$ is the Fourier transform of the external field at wave vector $\mathbf{q}$.
This theory is Gaussian, thus the partition function and free energy are easily
calculated. For a uniform magnetic field $h$, the free energy reads
$F_l(h) = \sum_{\mathbf{q}} \ln (\epsilon + \gamma^2 \mathbf{q}^2) - L_\parallel^2
\mu^2h^2/4\epsilon$. Here $\epsilon =\delta +\lambda$ as before, and $L_\parallel$ is the
linear size of the layer. The value of the Lagrange multiplier $\lambda$ follows from the
large-$N$ constraint
\begin{equation}
\langle \phi^2 \rangle = \frac 1 {L_\parallel^2} \frac {\partial F_l}{\partial \epsilon}
= \frac 1 {L_\parallel^2} \sum_{\mathbf{q}} \frac 1 {\epsilon + \gamma^2 \mathbf{q}^2}
+ \frac {\mu^2 h^2}{4\epsilon^2} =1~.
\label{eq:constraint_single}
\end{equation}
For small fields, $\mu h \ll \epsilon(h=0)$, the first term in the sum dominates, yielding
$\epsilon(h) = \epsilon(0) + O(h^2)$ with $\epsilon(0)$ given by (\ref{Eq:single_layer}).
In the opposite limit, $\mu h \gg \epsilon(h=0)$, the second term dominates, resulting in
$\epsilon(h)=\mu h/2$.

The magnetization of the single layer is easily computed by taking the appropriate derivative
of the free energy
\begin{equation}
m_l = - (1/L_\parallel^2) (\partial F_l / \partial h)_\epsilon =  \mu^2 h/2\epsilon~;
\label{eq:m_single}
\end{equation}
and the zero-field uniform susceptibility is given by
\begin{equation}
\chi_l =  \mu^2/2\epsilon(0)~.
\label{eq:chi_single}
\end{equation}
Other observables can be computed in an analogous fashion. For instance, the local
susceptibility $\chi_{l,loc}$ takes the same form as (\ref{eq:chi_single}) with $\mu^2$ replaced by
$\mu$.

\subsection{Critical point and weakly disordered Griffiths phase}
\label{Subsec:QCP+WD}

We now combine the single-layer observables with the strong-disorder renormalization
group results for the density (\ref{eq:n_scaling}) and
the moment (\ref{eq:mu_scaling})
of the surviving layers. In the present subsection we focus on
the critical point and the disordered Griffiths phase while the ordered Griffiths
phase will be addressed in Sec.\ \ref{Subsec:WO}.

The total magnetization in a magnetic field $h$ can be obtained by running the
renormalization group to the energy scale $\Omega_h = \mu_{\Omega_h} h$. All the surviving
layers have $\epsilon \ll \mu h$ and are thus fully polarized. The total magnetization
per site thus reads
\begin{eqnarray}
m(\delta,h) &=&  n_{\Omega_h}(\delta) \mu_{\Omega_h}(\delta) \nonumber\\
          &=&
          \left[\ln(\Omega_I/\Omega_h)\right]^{\phi-1/\psi}\Theta_{m}\left[\delta^{\nu\psi}
          \ln(\Omega_I/\Omega_h)\right]~.~~
\label{eq:m-scaling}
\end{eqnarray}
The scaling function is given by $\Theta_m(y) = X_n(y) X_\mu(y)$.
Now, using the fact that $\Omega_h = \mu_{\Omega_h} h$, we
find $m \sim [\ln(\Omega_I/h)]^{\phi - 1/\psi}$ (with double-logarithmic
corrections) at criticality, $\delta=0$. This implies that the critical isotherm
exponent $\bar \delta$ (commonly defined via $m \sim h^{1/\bar \delta}$) is formally infinite.
In the Griffiths phase, $\delta>0$, we obtain
$m \sim  h^{1/z} \delta^{\nu
+\nu\psi(1-\phi)(1+1/z)}\left[\ln(\Omega_I/h)\right]^{1+1/z}$.
As long as $z \sim \delta^{-\nu\psi}$ is larger than one (i.e., sufficiently close to the
critical point) this contribution dominates the regular linear-response term. We thus
find a non-universal power-law singularity in a finite temperature interval around the
critical point.

In the zero-field limit, the uniform susceptibility $\chi=\partial m /\partial h \sim h^{1/z-1}$
consequently diverges not just at the critical point but for all $z>1$, again
in an entire temperature range around the
critical point.
This result can also be obtained by summing (\ref{eq:chi_single})
over all layers using the spectral density $\rho(\epsilon)=dn_\Omega
/d\Omega|_{\Omega=\epsilon}$. In the Griffiths phase this gives the following
rare region contribution  to the susceptibility
\begin{equation}
\chi(h \to 0) \sim \left \{
\begin{array}{cc}
\infty & \quad (z>1) \\
\frac z {1-z} \Omega_I^{1/z-1} & \quad (z<1)
\end{array}
 \right. ~.
\label{eq:chi_griffiths}
\end{equation}

The specific heat can be obtained by summing the single-layer free energy $F_l$
over the spectral density $\rho(\epsilon)$ and taking the appropriate derivatives
with respect to the reduced temperature. As in the McCoy-Wu model, the resulting
specific heat is smooth across the transition.

\subsection{Weakly ordered Griffiths phase}
\label{Subsec:WO}

While the order parameter symmetry does not play a significant role on the disordered
side of the critical point where all conventional (non rare-region) excitations are
gapped, it becomes important on the ordered side of the transition where gapless
excitations exist even in the absence of our rare region physics. This leads to some
minor differences between our results and those of the McCoy-Wu model.

To determine the spontaneous magnetization on the ordered side of the transition,
we follow the strong-disorder renormalization group flow from $\Omega_I$ towards
$\Omega=0$. For small but nonzero $|\delta|$, i.e., close to the critical point,
the flow initially follows the critical trajectory until the renormalization group
length scale reaches the correlation length $\xi_\perp  \sim |\delta|^{-\nu}$. This occurs
at an energy $\Omega_\xi$ given by $\ln(\Omega_I/\Omega_\xi) \sim \xi_\perp^\psi \sim
|\delta|^{-\nu\psi}$. Beyond this scale, the system is essentially ordered, and
almost no layers will be removed under further action of the renormalization group.

We can therefore find the spontaneous magnetization by counting how many of the original layers
survive at length scale $\xi_\perp$. This leads to
\begin{equation}
m \sim n_{\Omega_\xi} \mu_{\Omega_\xi} \sim
[\ln(\Omega_I/\Omega_\xi)]^{\phi-1/\psi} \sim |\delta|^{\nu(1-\phi\psi)}~.
\label{eq:spontaneous_m}
\end{equation}
The order parameter critical exponent thus takes the value $\beta = \nu(1-\phi\psi)$.
In a small magnetic field $h$, the magnetization picks up a nonanalytic correction
which can be computed following the methods of Ref.\ \onlinecite{Fisher95}. We find
\begin{equation}
m(h)-m(0) \sim h^{1/(1+z)}~,
\label{eq:extra_m}
\end{equation}
implying that the (longitudinal) susceptibility $\chi=\partial m /\partial h \sim h^{-z/(z+1)}$
diverges in the zero-field limit everywhere in the weakly ordered Griffiths phase.
(The transverse susceptibility is infinite everywhere in the ordered phase simply because
of the continuous order parameter symmetry.)

Another important property of the ordered phase of a continuous symmetry magnet is the spin-wave
stiffness which can be defined via the change of the free energy with a twist in the
boundary conditions. In our system, we must distinguish the parallel spin-wave stiffness
from the perpendicular one. To find the parallel spin-wave stiffness $\rho_{s}^{\parallel}$,
we apply boundary conditions at $x=0$ and $x=L_\parallel$ such that the spins at the two
ends are at a relative angle $\Theta$. In the limit of small $\Theta$ and large $L_\parallel$,
the free energy density $f$ depends on $\Theta$ as
\begin{equation}
f(\Theta) -f(0) = \frac 1 2   \rho_s^{\parallel}
\left( \frac {\Theta}{L_\parallel}\right)^2
\label{eq:rho_s^parallel_def}
\end{equation}
which defines $\rho_s^{\parallel}$.

In our system, the free energy cost due to the twist is simply the sum over all
layers participating in the long-range order. Each layer has the same twisted
boundary conditions and the perpendicular bonds (which are not twisted) do not
contribute. The bare stiffness of a single layer is given by $\gamma^2$. Because
$\gamma^2$ is additive under the strong-disorder renormalization group,
$\rho_s^\parallel$ behaves like the layer moment per site, $\rho_s^\parallel \sim \mu$.
The calculation of the total parallel spin-wave stiffness thus proceeds analogously
to the total magnetization yielding
\begin{equation}
\rho_s^\parallel \sim \gamma_0^2 \, |\delta|^\beta = \gamma_0^2 \,
|\delta|^{\nu(1-\phi\psi)}~.
\label{eq:rho_s^parallel_result}
\end{equation}

If a global twist is applied perpendicular to the layers, i.e., between the bottom ($z=0$) and
top ($z=L_\perp$) of the stack, the local twist $\Theta_z$ between layers $z$ and $z+1$
will vary from layer to layer according to the local $J_z^\perp$.
The total free energy cost can be written as
\begin{equation}
f(\Theta) - f(0) \sim \frac 1 {2 L_\perp} \sum_z \rho_z \Theta_z^2
\label{eq:rho_s^perp_def}
\end{equation}
with $\rho_z \sim J_z^\perp$. Minimizing $f(\Theta)-f(0)$ under the constraint
$\sum_z \Theta_z = \Theta$ gives $\Theta_z \sim 1/\rho_z$ and
\begin{equation}
f(\Theta) - f(0) \sim \left(L_\perp \sum_z \rho_z^{-1} \right)^{-1}~.
\label{eq:rho_s^perp_result}
\end{equation}
To obtain an upper bound for $f(\Theta) - f(0)$, we estimate $\sum_z \rho_z^{-1}$ by its
largest contribution, $\rho_{min}^{-1} \sim (J_{min}^\perp)^{-1}$. In the weakly ordered
Griffiths phase, the fixed-point distribution of $J^\perp$ is gapless,\cite{Fisher95} and
$J^\perp_{min}$ vanishes as $L_\perp^{-z}$ in the thermodynamic limit $L_\perp \to
\infty$. We conclude $f(\Theta) - f(0) \sim L_\perp^{-z-1}$, implying that the global
perpendicular stiffness vanishes, $\rho_s^{\perp}=0$ (for $z>1$). The weakly ordered
Griffiths phase is thus very peculiar because the system displays long-range
ferromagnetic order but it has no (perpendicular) spin-wave stiffness.

\section{Finite-size effects}
\label{Sec:FSE}

The results in Sec.\ \ref{Sec:TD} were for an infinite system (in the thermodynamic
limit). Here, we briefly discuss the effects of a finite system size in either
parallel or perpendicular direction.

We start with a finite in-plane (parallel) size $L_\parallel$. It plays the same role
as a finite temperature in the quantum phase transitions in Refs.\
\onlinecite{Fisher92,Fisher95,HoyosKotabageVojta07,VojtaKotabageHoyos09} where the
inverse temperature is the system size in imaginary time direction. Solving the
large-$N$ constraint for a single layer of linear size $L_\parallel$ gives
$\epsilon(L_{\parallel}) = \epsilon(\infty) + O(1/L_{\parallel}^2)$ for
$\epsilon(\infty) \gg 1/L_\parallel^2$. Here, $\epsilon(\infty)$ is the thermodynamic
limit result given in (\ref{Eq:single_layer}). In the opposite limit,
$\epsilon(\infty) \ll 1/L_\parallel^2$, we obtain $\epsilon(L_\parallel)
=1/L_\parallel^2$. Thus, a finite $L_\parallel$ cuts off the low-$\epsilon$ tail
in the spectral density $\rho(\epsilon)$.

As an example of the resulting finite-size effects in thermodynamic quantities we
now discuss the dependence of the susceptibility on $L_\parallel$. Within the
strong-disorder renormalization group, it can be found by running the
renormalization group to the scale $\Omega_L= 1/L_\parallel^2$. Beyond that scale,
$\epsilon$ is not renormalized further down. All surviving layers now have
$\epsilon \gg J^\perp$ and can thus be treated as independent. Using (\ref{eq:chi_single}),
the uniform susceptibility
of a system of size $L_\parallel$ is consequently given as the sum over all layers
surviving at scale $\Omega_L$,
\begin{equation}
\chi(\delta,L_\parallel) = n_{\Omega_L}(\delta) \mu_{\Omega_L}^2(\delta) /2\Omega_L~.
\label{eq:FS_chi}
\end{equation}
At criticality, $\delta=0$, this leads to $\chi \sim L_\parallel^2 \,
[\ln(L_\parallel/a)]^{2\phi-1/\psi}$. We emphasize that $\chi$ is the susceptibility
\emph{per volume}, so $L_\parallel^2$ is not simply a geometric factor but indicates the
divergence of the susceptibility in the thermodynamic limit. In the weakly disordered
Griffiths phase, the same calculation gives (up to logarithmic corrections)
a non-universal power-law dependence,
$\chi \sim \delta^{\nu+2\nu\psi(1-\phi)} L_\parallel^{2-2/z}$. In the weakly ordered
Griffiths phase, we need to take into account that long-range order is not possible for
any finite $L_\parallel$. Thus all layers surviving at scale $\Omega_L$ will again
contribute to the susceptibility. In contrast to the weakly disordered Griffiths phase,
the typical moment of a layer is proportional to its thickness $\mu_{\Omega_L} \sim m_0
n^{-1}_{\Omega_L}$ where $m_0$ is the bulk magnetization. In the weakly ordered
Griffiths phase, we thus obtain $\chi \sim \delta^{\nu-\nu\phi\psi} L_\parallel^{2+2/z}$.
All of our results for the $L_\parallel$ dependence of the uniform susceptibility are
completely analogous to the corresponding temperature dependencies at the quantum phase
transition in the random transverse-field Ising chain in Ref.\ \onlinecite{Fisher95}.
They are also compatible with finite-size scaling using that $1/L_\parallel^2$ scales
like $\epsilon$ (or, equivalently, like a magnetic field $H$).
Other observables can be worked out in a similar fashion.

We now turn to the effects of a finite size $L_\perp$ in perpendicular direction, i.e.,
the effects of a finite number of layers in our stack. We expect these effects to be
particularly important experimentally because growing samples containing a macroscopic
number of layers will often be difficult. The origin of finite-size effects in $L_\perp$
is that finite-size samples do not contain rare regions (strongly-coupled layers)
beyond a certain thickness or, equivalently, they do not contain rare regions with
$\epsilon < \Omega_{min} (L_\perp)$.

Within the strong-disorder renormalization group, the relation between system size and
the cutoff energy scale $\Omega_{min} (L_\perp)$ can be worked out using the density of
surviving layers $n_\Omega$. In a typical sample of size $L_{\perp}$, the \emph{number}
of layers surviving at renormalization group scale $\Omega$ is given by
$L_{\perp} \, n_\Omega$. The cutoff scale is thus defined by $L_{\perp} \,
n_{\Omega_{min}}=1$. At criticality, this implies
\begin{equation}
\ln(\Omega_I/\Omega_{min}) \sim L_\perp^\psi
\label{eq:Omega_min_cp}
\end{equation}
reflecting the activated character of finite-size scaling in perpendicular direction. In
the two Griffiths phases, we obtain
\begin{equation}
\Omega_{min} \sim |\delta|^{-\nu z} L_\perp^{-z}~.
\label{eq:Omega_min_griffiths}
\end{equation}

As the first example of the resulting finite-size effects in thermodynamic quantities, we
consider the magnetization-field curve $m(h)$.
To do so, we compare the field-induced renormalization group
cutoff $\Omega_h=\mu_{\Omega_h} h$ and the finite-size cutoff $\Omega_{min}$. As long as
$\Omega_h > \Omega_{min}$, the finite system size $L_\perp$ has only a negligible effect
on the magnetization. However, $L_\perp$, cuts off the nonlinear low-field tail of $m(h)$
once $\Omega_h < \Omega_{min}$. At criticality, this happens for fields below
$h_{min}$ given by $\ln(\Omega_I/h_{min}) \sim L_\perp^\psi$. In the weakly disordered
Griffiths phase, the nonlinear $m(h)$ curve is cut off below
$h_{min} \sim \delta^{-\nu z -\nu \psi (1-\phi)} L_\perp^{-z}$. In the weakly ordered
Griffiths phase, the calculation is slightly more involved because we first need to
resolve the relation between $\Omega_h = \mu_{\Omega_h} h$ using $\mu_{\Omega_h} \sim m_0
n^{-1}_{\Omega_h}$ where $m_0$ is the bulk magnetization. We finally obtain
$h_{min} \sim |\delta|^{\nu \phi \psi - \nu(1+z)} L_{\perp}^{-(1+z)}$.

The zero-field susceptibility of a typical sample of perpendicular size $L_\perp$ can be calculated
by summing (\ref{eq:chi_single}) over all layers with the spectral density
$\rho(\epsilon)$ cut off at $\epsilon = \Omega_{min}$. Alternatively, it can be estimated
by $(\partial m /\partial h)_{h_{min}}$. At criticality, the susceptibility diverges
exponentially with system size,
$\chi \sim \exp(A L_\perp^\psi)$ with $A$ a constant. In the weakly disordered Griffiths
phase, we find a non-universal power law,
$\chi \sim \delta^{1+\nu z + 2 \nu \psi (1-\phi)} L_\perp^{z-1}$.

Finally, we discuss the finite-size behavior of the perpendicular spin-wave stiffness
$\rho_s^\perp$ in the weakly ordered Griffiths phase. In Sec.\ \ref{Subsec:WO}, we showed
that $\rho_s^\perp$ vanishes in the thermodynamic limit, because the fixed-point
distribution of $J^\perp$ is gapless. A finite perpendicular size $L_\perp$ establishes a
lower bound for  $J^\perp$ in a typical sample. From (\ref{eq:Omega_min_griffiths}) we
obtain $J^\perp_{min} \sim L_\perp^{-z}$. Thus, the perpendicular spin-wave stiffness of
a typical sample vanishes as $\rho_s^\perp \sim L_\perp^{1-z}$ with increasing system
size (for $z>1$).

\section{Critical dynamics}
\label{Sec:Dynamics}

It is well known that dynamic critical phenomena
show stronger rare region effects and Griffiths singularities than the corresponding
thermodynamic critical phenomena at classical phase transitions.
 In particular, rare regions dominate the long-time
dynamics in a conventional classical Griffiths phase
\cite{Dhar83,RanderiaSethnaPalmer85,Bray88,Bray89} even though they provide only
small corrections to the thermodynamics. In this section, we therefore study the critical
dynamics in our randomly layered Heisenberg magnet.

The classical Heisenberg model does not have any internal dynamics, we therefore add
a phenomenological dynamics to our system. Here, we focus on the simplest case,
a purely relaxational dynamics corresponding to model A in the classification of Hohenberg
and Halperin.\cite{HohenbergHalperin77} Microscopically,
this type of dynamics can be realized, e.g., via the Glauber \cite{Glauber63} or
Metropolis \cite{MRRT53} algorithms. Other dynamical algorithms can be studied using
similar methods (including model J which describes the dynamics of real Heisenberg spins).
This remains a task for the future.

To characterize the dynamic critical behavior, we calculate the average autocorrelation
function
\begin{equation}
C(t) = \frac 1 {L_\perp L_\parallel^2}  \int d^3 r \langle
\phi(\mathbf{r},t)\phi(\mathbf{r},0)\rangle~,
\label{eq:autocorr_def}
\end{equation}
where $\phi(\mathbf{r},t)$ is the order parameter at position $\mathbf{r}$ and time $t$.
In addition, we also determine the dynamic susceptibility $\chi(\omega)$.

Let us begin by considering the dynamics of a single layer with moment $\mu$ per site and a
renormalized local distance $\epsilon$ from criticality. Because a single layer cannot display long-range
order, the correlations decay exponentially in time. The dependence of the correlation
(relaxation) time $\xi_t$ on $\epsilon$ can be found following the heuristic arguments
of Bray.\cite{Bray88} He considered a correlation volume $\xi_\parallel^2 \sim
1/\epsilon$ which he assumed to be in the magnetic state with total magnetization
$M_0 \sim \mu \xi_\parallel^2 \sim \mu/\epsilon$. The relaxation of the magnetization
occurs mainly via diffusion of the order parameter vector on a sphere of radius $|M_0|$
due to thermal noise (because there are no energy barriers in the case of Heisenberg
symmetry). The noise at different points in space and time adds incoherently. Thus, according
to the central limit theorem, the change in magnetization after time $t$ is
$\Delta M(t) \sim t^{1/2} (\mu/\epsilon)^{1/2}$. Defining $\xi_t$ as the time when
$\delta M(\xi_t) \sim M_0$, we obtain
\begin{equation}
\xi_t(\epsilon) \sim \mu/\epsilon~.
\label{eq:xi_t}
\end{equation}
At criticality and in the weakly disordered Griffiths phase, $\mu$ only provides logarithmic
corrections to the leading $1/\epsilon$ dependence.

The same result can also be obtained more formally from the single-layer
Langevin equation
\begin{equation}
\frac {\partial \phi(\mathbf{q},t)}{\partial t} = -2 \Gamma_0 (\epsilon + \gamma^2
\mathbf{q}^2) \phi(\mathbf{q},t) + \Gamma_0 \mu h(\mathbf{q},t) +\eta(\mathbf{q},t)
\label{eq:Langevin}
\end{equation}
where $h(\mathbf{q},t)$ is a time-dependent magnetic field, $\eta(\mathbf{q},t)$ is the usual
$\delta$-correlated noise and $\Gamma_0$ fixes the overall time scale. To find the
autocorrelation function of a single layer, we solve (\ref{eq:Langevin}) for $h(\mathbf{q},t)=0$ and
insert the solution into (\ref{eq:autocorr_def}). In the asymptotic long-time limit, $\Gamma_0\epsilon
t \gg 1$, we
find
\begin{equation}
C_l(t) \sim \exp(-2 \Gamma_0 \epsilon t)/ (\Gamma_0 \epsilon t)~,
 \label{eq:C_l}
\end{equation}
in agreement with
the heuristic estimate (\ref{eq:xi_t}). Solving the Langevin equation in the presence of
a field allows us to calculate the single-layer dynamic susceptibility
$\chi_l(\mathbf{q},\omega)=\partial m(\mathbf{q},\omega) /\partial h(\mathbf{q},\omega)$.
For a uniform field, $\mathbf{q}=0$, this results in
\begin{equation}
\chi_l(\omega) = \mu^2 /(2\epsilon -i\omega/\Gamma_0)~.
\label{eq:chi_eps}
\end{equation}

After having discussed the single-layer dynamics, we now turn to the full system.
To find the average autocorrelation function at time $t$,
we run the strong-disorder renormalization
group to the scale $\Omega_t = 1/t$. All layers eliminated during this procedure have
correlation times $\xi_t \ll t$ and do not contribute to the autocorrelation function.
Surviving layers have $\xi_t \gg t$, they thus contribute proportional to
their moment $\mu$ per site, giving $C(t) \sim n_{\Omega_t} \mu_{\Omega_t}$. At
criticality, this leads to an ultraslow logarithmic decay of the autocorrelation
function,
\begin{equation}
C(t) \sim [\ln(t/t_0)]^{\phi-1/\psi}
\label{eq:C_CP}
\end{equation}
with $t_0$ a microscopic time scale. In the weakly disordered Griffiths phase, the same calculation
yields, up to logarithmic corrections, a
non-universal power law
\begin{equation}
C(t) \sim \delta^{\nu+\nu\psi(1-\phi)} t^{-1/z}~.
\label{eq:C_Griffiths}
\end{equation}
The same time dependence  also follows from averaging (\ref{eq:C_l}) over the spectral density
$\rho(\epsilon)$. The power-law decay (\ref{eq:C_Griffiths}) is much slower than the
stretched exponential found in conventional classical Griffiths
phases.\cite{Dhar83,RanderiaSethnaPalmer85,Bray88,Bray89}
Interestingly, (\ref{eq:C_CP}) and (\ref{eq:C_Griffiths}) are reminiscent of the behavior
at certain classical \emph{non-equilibrium} phase transitions with
disorder.\cite{Noest86,HooyberghsIgloiVanderZande03,VojtaDickison05}

The uniform dynamic susceptibility can be computed in an analogous manner. At criticality, we
find its imaginary part to behave as
\begin{equation}
\chi{''}(\omega) \sim \frac 1 \omega [\ln(\Gamma_0/\omega)]^{2\phi-1/\psi}~.
\label{eq:chi_CP}
\end{equation}
In the weakly disordered Griffiths phase, we again obtain a power law,
\begin{equation}
\chi{''}(\omega) \sim \delta^{\nu + 2\nu\psi(1-\phi)} \omega^{1/z-1}~.
\label{eq:chi_Griffiths}
\end{equation}
For the local dynamic susceptibility $\chi_{loc}(\omega)$, the corresponding relations are
$(1/\omega) [\ln(\Gamma_0/\omega)]^{\phi-1/\psi}$ and
$\delta^{\nu + \nu\psi(1-\phi)} \omega^{1/z-1}$ at criticality and in the
Griffiths phase, respectively.

At first glance, the above results for $C(t)$ and $\chi{''}_{loc}(\omega)$ appear to violate
the fluctuation dissipation theorem which requires $\chi{''}_{loc}(\omega) =
(\omega/2T) C(\omega)$  where $C(\omega)$ is the Fourier transform of the
autocorrelation function. The reason for the disagreement is that the relaxation
time (\ref{eq:xi_t}) diverges for the largest rare regions (which correspond to effective
layers with the smallest $\epsilon$). Thus, the layers that dominate the long-time tail
of $C(t)$ are not in equilibrium, and the fluctuation-dissipation theorem is not applicable.
Technically, the disagreement is caused by the fact that (\ref{eq:C_l}) cannot be
used for the layers with the smallest $\epsilon$ at any finite time.

\section{Discussion and conclusions}
\label{Sec:Conclusions}

In summary, we have investigated the phase transition in a three-dimensional randomly
layered classical Heisenberg magnet. We have employed a strong-disorder renormalization
group to show that the critical point is of unconventional infinite-randomness character.
Somewhat surprisingly, the critical behavior can be found exactly, making our system
one of the very few examples of three-dimensional systems with exactly known critical
exponents. The critical point is accompanied by strong power-law Griffiths singularities
(which are often called \emph{quantum} Griffiths singularities because they generically occur in
quantum systems but not in classical systems).
In addition to the thermodynamics, we have also studied the critical dynamics
within model A of the Hohenberg-Halperin classification. It is characterized by an
ultraslow relaxation of the magnetic correlations at criticality as well as in the
Griffiths phase.

Our findings can be related to a broader
classification\cite{VojtaSchmalian05,Vojta06} of phase
transitions with quenched disorder. This classification is based on the
effective dimensionality of the defects or, equivalently,
the rare regions. Three classes can be distinguished:
(i) If the defect dimensionality is below the lower critical
dimension $d_c^-$ of the problem, the resulting critical point is conventional,
and the Griffiths singularities are exponentially weak.
(ii) If the defect dimensionality is exactly equal to the lower critical
dimension, the critical point is of infinite-randomness type and
accompanied by power-law ``quantum'' Griffiths singularities.
(iii) Finally, if the defects are above the lower critical dimension,
individual regions can order independently, leading to a smeared transition.

The randomly layered Heisenberg magnet falls into class (ii) because the dimensionality
of the planar defects is two, identical to the lower critical dimension of the classical
Heisenberg model. The results of this paper are therefore in complete agreement with the
general classification. It is worth noting, that the behavior of a randomly layered
\emph{Ising} magnet is very different. The lower critical dimension of a classical Ising
model is one, thus planar defects are above the lower critical dimension. Consequently,
the phase transition in a randomly layered Ising magnet is smeared by the
disorder.\cite{Vojta03b,SknepnekVojta04}

We emphasize that the above classification also helps resolve the puzzling question
posed at the end of Sec.\ \ref{Subsec:Recursions}, \emph{viz.}, why systems as different
as the randomly layered Heisenberg magnet and the McCoy-Wu model (or, equivalently, the
random transverse-field Ising chain) end up in the same universality class. The crucial
point is that even though these two systems have different order parameter symmetries and
dimensionalities, the defects are exactly at the lower critical dimension in both cases:
$d_c^-=2$ for the classical Heisenberg model and $d_c^-=1$ for the Ising model. These arguments
demonstrate why both the McCoy-Wu model and our randomly layered Heisenberg
model end up having infinite-randomness critical points and thus the same scaling
scenario;
they do not yet explain why the two systems share the same critical exponent values.
The agreement of the exponent \emph{values} follows from the fact that both systems are
random in one direction which leads to coarse graining in one dimension within the
strong-disorder renormalization group. Moreover, the renormalization group fixed point
only depends on the multiplicative structure the recursion relations (\ref{eq:J-tilde})
and (\ref{eq:e-tilde}) and not on model-dependent prefactors.

Our explicit calculations have been performed in the large-$N$ limit of the
order parameter field theory (\ref{Eq:LGW}). However, the critical fixed point
stays valid for all $N > 2$ including the case of Heisenberg symmetry,
$N=3$. To see this, we need to confirm that the recursion
relations (\ref{eq:J-tilde}) and (\ref{eq:e-tilde}) remain unchanged
for any $N>2$. The
multiplicative structure
of the recursion (\ref{eq:J-tilde}) for $J^\perp$ follows directly
from second order perturbation theory and is thus the same for all $N$.
In contrast, the multiplicative structure of the recursion relation (\ref{eq:e-tilde})
for the renormalized distance $\epsilon$ from criticality follows from the fact
that a single layer of a continuous symmetry order parameter ($N>2$) is exactly
at the lower critical dimension. This implies an exponential dependence of $\epsilon$
on the moment of the effective layer and thus the multiplicative form of (\ref{eq:e-tilde});
for details see the corresponding discussion in
Ref.\ \onlinecite{VojtaKotabageHoyos09}. Consequently, up to unimportant prefactors,
both recursion relations remain valid for any $N>2$ and with them the infinite-randomness
critical point scenario found in this paper.

The strong-disorder renormalization group allowed us to identify the
infinite-randomness fixed point and verify its stability. However, it cannot tell
whether or not a weakly or moderately disordered system will flow toward
this fixed point. This is due to the fact that for weak disorder, the renormalization group
recursions are not very accurate. To gain further insight, it is useful to look
at the behavior in the weak disorder limit. The effects of weak disorder
on a clean critical point are governed by the Harris
criterion\cite{Harris74} that states that the
clean fixed point is stable against disorder, if its correlation length
exponent $\nu$ fulfills the inequality $d_r \nu >2$ where $d_r$ is the number of
dimensions in which there is randomness. In our case, $d_r=1$ and the
correlation length exponent of the clean 3D Heisenberg model is\cite{CHPRV02}
$\nu\approx 0.711$.
Therefore, the clean 3D Heisenberg critical point violates the inequality
$d_r \nu >2$ implying that it is unstable against weak planar disorder.
(It is also unstable against linear disorder, $d_r=2$
(see Ref.\ \onlinecite{SknepnekVojtaVojta04}), but stable against
the usual point disorder, $d_r=3$.)
Within a renormalization group approach this means that weak
planar disorder initially increases under renormalization suggesting that our fixed point
may control the critical behavior for all bare disorder strength. A more complete
answer to this question will likely come from experiment and computer simulations.

Experimental verifications of infinite-randomness critical behavior and the
accompanying power-law ``quantum'' Griffiths singularities have been hard to come
by, in particular in higher-dimensional systems. Only very recently, promising
measurements have been reported\cite{Westerkampetal09,UbaidKassisVojtaSchroeder10}
of the quantum phase transitions in CePd$_{1-x}$Rh$_x$ and Ni$_{1-x}$V$_x$.
We hope that our work opens an alternative avenue to observe these phenomena
in systems that may be easier to study experimentally.\\

\section*{Acknowledgements}

This work has been supported in part by the NSF under grant nos. DMR-0339147
and DMR-0906566 and by Research Corporation. T.V. acknowledges the hospitality of
the Aspen Center of Physics as well as IIT Madras where parts of the research have
been performed.

\bibliographystyle{apsrev}
\bibliography{../00Bibtex/rareregions}
\end{document}